\documentclass[prl,showpacs]{revtex4}
\usepackage{graphics}
\usepackage{epsfig}
\usepackage{graphicx}

\begin{document}
\title{When Do Superfluidity and Long Range Order Imply Entanglement?}
\author{Vlatko Vedral}
\affiliation{School of Physics and Astronomy, University of Leeds, United Kingdom and\\National University of Singapore, Singapore}

\pacs{03.75.Gg, 03.67.Mn}
\date{\today}

\begin{abstract}
We investigate tacitly assumed relationships between the concepts of super-fluidity (-conductivity), long range order and entanglement. We prove that the three are by no means equivalent, but that notwithstanding, some rigorous implication can be established between them. This leads to three different, albeit frequently related, notions of ``criticality", all of which are exemplified within the Hubbard model in the low density regime. We use Peierls' method of twisted Hamiltonians to link the existence of entanglement to superfluidity and (quasi)-long range order. As an application of our formalism, we show that recent experiments with cold atoms already prove the existence of the field theoretic, spatial entanglement in two dimensions. More interestingly, the appearance of entanglement in these experiments seems to be intimately related to the phase transition of the Kosterlitz Thouless type.   
\end{abstract}

\maketitle

\section{Introduction} 

Phase transitions play a very important role in physics because they challenge us to describe these inherently macroscopic phenomena in terms of the underlying, microscopic, mechanisms \cite{Yeomans}. We have many different types of phase transitions, ranging from a solid to liquid to gas transition of large numbers of atoms, to the quantum mechanical electron (Cooper) pairing signifying the onset of superconductivity. All different phase transitions, however, have one crucial aspect in common. Simply put, by varying external conditions a certain same lump of matter (or light) can fundamentally change its macroscopic behaviour. Ehrenfest was the first to classify phase transitions according to which macroscopic property in particular suffers this abrupt change. If the first derivatives of the free energy (such as the internal energy, or magnetisation) are the observables in question, then the phase transition is of the first order; if the second derivatives become discontinuous (e.g. magnetic susceptibility or heat capacity), then the phase transition is of the second order, and so on. In spite of this advance, a coherent account of phase transitions was still amiss until the application of quantum mechanics to many body physics.    

The first theory of superfluid critical phenomena was developed by Landau \cite{Landau} and also goes under the name of mean-field theory. It suggests that any phase transition can be viewed as an order to disorder transition. To quantify it we therefore need to identify a function (defined at all points in space), called an order parameter, whose non-zero value identifies that we are below the critical point (e.g. temperature) where order exists. Above the critical temperature the order parameter disappears. The existence of a function that stretches across the whole physical system in the ordered phase immediately suggests that different parts of the system, even if far apart, are in some sense correlated through this order parameter function. This gives rise to the notion of long range order (LRO), introduced by Penrose and Onsegar \cite{Penrose}, in order to understand Bose condensation. The condition for the existence of LRO is given by
\begin{equation}
\lim_{_{|x-x'|\rightarrow \infty}} \langle \Psi^{\dagger} (x) \Psi (x') \rangle \Longrightarrow const>0
\end{equation}  
and for Penrose and Onsegar is identical to the system being in a Bose condensate. 
The constant to which the two point correlations tend in the large size of the system limit is, in fact, the order parameter discussed before. For us here the interesting question will be if this long range order implies the existance of entanglement and if so, what kind of entanglement might that be. Before that let us first finalise our brief exposition of phase transitions. 

By the time low temperature superconductivity was properly described by Bardeen, Cooper and Schriefer \cite{BCS}, it was clear that the LRO does not quite apply to it immediately. Electrons in a superconductor are not correlated across such large distances that they cover the whole of the superconducting sample. Yang was the first to realise that we need a new notion of order which he termed the ``off-diagonal long range order" (ODLRO) \cite{Yang}. If instead of $\Psi(x)$ representing a single electron at position $x$, this operator represented a Cooper pair at the same $x$, 
$\Psi (x) = \psi_{\uparrow} (x) \psi_{\downarrow} (x)$
then the Penrose Onsegar condition for LRO would still be satisfied by this new operator. The proper order parameter is therefore the wave function for a Cooper pair - a pair of electrons in a spin singlet state. 
We can then think of a superconductor as a condensate of Cooper pairs, though electrons are still fermions and behave as such. (In BCS theory of superconductivity, we can likewise think of the energy gap between the ground state with Cooper pairs and the excited states where they start to break up as a good order parameter. The gap disappears exactly above the critical temperature, when the superconductor transforms into an ordinary conductor. Both of these formulations are equivalent and lead to the same calcualtion of the critical temperature for superconductivity.)

The next big surprise in understanding phase transition came in the seventies when it was discovered that we can actually have a phase transition without the existence of any long range order. This discover was, interestingly, predated by a number of different proofs of the fact that continuous phase transitions cannot exist in 1 and 2 spatial dimensions \cite{Peierls, Mermin, Hohenberg}. The reason for this is very simple and it lies in the fact that any long range order is destroyed by thermal fluctuations in low dimensions. However, what can happen is that a kind of short range order emerges in two dimensions (we will refer to this a quasi long range order) out of a disordered phase. This means that below some temperature the correlations exhibit a polynomial drop with the distance like so
$\lim_{|x-x'|\rightarrow \infty}\langle \Psi^{\dagger} (x) \Psi (x') \rangle \longrightarrow 1/|x-x'|^p$, 
where $p$ is some power. Above this temperature the drop is typically exponential. The mechanism for this transition was explained by Berezinskii and Kosterlitz and Thouless \cite{BKT} (BKT) and has been experimentally verified a number of times since. In summary, therefore, we can have long range order, off diagonal long range order as well as quasi long range order present in matter and all linked to phase transitions.

The whole story about order-disorder transitions can and has been applied to quantum phase transitions as well \cite{Sachdev}. These transitions occur at zero temperature (in practice, at low temperatures) and are driven by changes of some parameter other than the temperature, such as the external magnetic field in a spin chain or a the doping parameter in a high temperature superconductor. Quantum phases can also exhibit long range order and quasi long range order. The point here, of course, is that any correlations are now likely to imply some type of entanglement, since we have an overall pure state\cite{OAFF}. In much the same way that different correlations in three dimensions lead to different crystals, we might expect that different types of entangled state will lead to different phenomena in quantum critical regions. 

\section{Independence of entanglement and quantum order} 

In this paper we address the issue of if and how quantum criticality and entanglement are related. We will find that the relationship is not entirely straightforward, but that some concrete conclusions can nevertheless be drawn. 
Before we start our more general discussion, we note that entanglement and quantum ODL are seemingly not 
strictly related to one another: each can exist without the presence of the other. For example, a product of coherent states of equal amplitude but at different positions $x$, $\prod_x |\alpha (x)\rangle$,
is spatially a disentangled state (by construction), but it exhibits ODLRO (equal to $|\alpha|^2$). On the other hand, the state of the GHZ type 
$|000..0\rangle + |111..1\rangle$ 
is clearly entangled, but LRO does not exist. The latter can be seen by the fact that the operator combination $\psi^{\dagger}\psi$ vanishes for states with the same number of excitations such as $|00\rangle$ and $|11\rangle$ (Here it is more appropriate to introduce Pauli raising and lowering operators, but we leave this for later).  Therefore, any kind of criticality that is indicated by LRO can be very different from the criticality signified by the vanishing of entanglement. Similar statements can be made about entanglement and quasi long range order. Entanglement is, in fact, just a more complex form of quantum order, and, just like correlations, it can exist between any collection of spins, be they close to each other or on different sides of the investigated system (for a recent review of entanglement in many-body systems see \cite{Amico}). 

If, as above, we can show that entanglement and LRO are independent, why has there been so much work on entanglement and phase transitions? The simple answer is that states do not encode all the information about the actual physics of the system. The Hamiltonian plays an equally important role. It can easily happen that the states we discussed above (coherent and GHZ) are never the eigenstates of the relevant Hamiltonian. And this is exactly what will happen in the situations analysed below. We will show that entanglement is, in spite of the above arguments to the contrary, intimately connected to quantum order and critical phenomena.     

Our approach here will entirely be based on physically observable effects. The notions of different order parameters are useful precisely because they are ultimately linked to some underlying physics. Long range order in superconductors, for example, has been shown to imply the Meissner effect as well as the notion of flux quantisation \cite{flux} (in superfluids it leads to irrotational flow and quantized vortices \cite{vortex}). It is in this sense that it would be desirable to view entanglement: can we say that superconductivity (-fluidity), or some other critical phenomena, are in any way dependent on entanglement? Here we analyse this question in detail and show that the answer is affirmative. Our example system will be very simple, but it contains all the necessary elements to draw some general conclusions. We will analyse both classical (i.e. $T>0$) and quantum ($T=0$) criticality and show that recent experiments with cold atoms already confirm the existence of field theoretic, particle number, entanglement.  

\section{Analysis of spinless Hubbard, XX model} 

In order to remain close to physics, we consider the so called Bose -Hubbard model. This model is used to investigate various superconducting and superfluid behaviours, ranging from High $T_c$ superconductors to cold atom gases in optical lattices. The model has quite a complex phase diagram \cite{Bose-Hubbard}, but we will only be interested in some special regimes relevant to our study of entanglement. 
The Bose-Hubbard Hamiltonian is given by
\begin{equation}
H_{BH} = J \sum_i (b^{\dagger}_i b_{i+1} + b_i b^{\dagger}_{i+1}) +\frac{U}{2}\sum_i (n_i-1)n_i \; \; , 
\end{equation}
where the first term describes the nearest-neighbour site hopping of bosons, ($b^{\dagger}$ and $b_{i+1}$ are the usual bosonic raising and lowering operators respectively), and the second term is the on site repulsion between bosons. In the limit of low density, the $n_i^2$ term can be ignored and we obtain the so called $XX$ Hamiltonian:
\begin{equation}
H = -J \sum_i \sigma^x_i\sigma^x_{i+1} + \sigma^x_i\sigma^x_{i+1} - \mu \sum_i \sigma^z_i 
\end{equation}
where $\sigma^x  =  b + b^{\dagger}$, $\sigma^y  =  i(b - b^{\dagger})$ and
$\sigma^z =  1 - 2 b^{\dagger}b$ are the usual Pauli matrices and we can think of $\mu$ as the chemical potential \cite{Sachdev}. They represent two level systems where the states $|0\rangle$ and $|1\rangle$ are the boson occupation numbers at each site (an empty site and one boson in the site respectively \cite{V}). Physically this should be clear, since by making the density very low, we preclude more than one particle from occupying each of the sites. Note that this Hamiltonian would also be obtained if we investigate the spinless fermion Hubbard model since fermions obey the Pauli exclusion principle and therefore we cannot have more than one electron per site.  

The $XX$ model has been extensively studied and its spectrum is well understood through the Jordan-Wigner transformation \cite{Sachdev}. However, since we would like to understand superfluidity in the original Hubbard model we need to know the response of the system with such a Hamiltonian to introducing external perturbations (here we study a one dimensional model; two dimensions are needed for proper superfulidity, but the one dimensional discussion leads to the same conclusion - see the Appendix for a discussion of this point). In order to study criticality in the $XX$ model we introduce the so called ``twisted" Hamiltonian \cite{Shastry,Shastry2}:
\begin{equation}
H_{\theta} = -J \sum_i e^{i\theta} \sigma^+_i\sigma^-_{i+1} + \sigma^-_i\sigma^+_{i+1}e^{-i\theta}  - \mu \sum_i \sigma^z_i 
\end{equation}
obtained by imposing Peierls phases on each Pauli raising and lowering operator:
$\sigma^-  \rightarrow  e^{i\theta} \sigma^-$ and $\sigma^+  =   e^{-i\theta}\sigma^+$.
We can think of phases arising from an imposition of an external field (like, for example, the Aharonov-Bohm effect, where the charge encirculating the vector potential, $A$, gains the phase $e^{-i\int Adl}$). The twisted Hamiltonian therefore represents the response of the system to an external disturbance and this is what gives rise to the physics of various critical phenomena. The superfluid density can now be obtained from the assumption that the difference between the normal and ``twisted" energy is - to the lowest order - given by the superfluid kinetic energy. So, when we externally weakly perturb the system, its response is to move its superfluid component (if there is such a component). If the response is null to the lowest order, our material is not in its critical phase. Translated into mathematics our statement reads \cite{Burnett}:
\begin{equation}
\langle H_{\theta} \rangle - \langle H \rangle = \frac{1}{2} Nf_s m v_s^2
\end{equation}
where $N$ is the number of particles in the superfluid, $m$ is the mass of each particle and $v_s$ the superfluid velocity. The superfluid density is denoted by $f_s$. The superfluid velocity is given by 
$v_s = \nabla \theta$,
where $\theta$ is the phase of the superfluid macroscopic wave function - the order parameter (and is the same angle that enters the twisted Hamiltonian). We will assume that the superfluid phase varies linearly with distance, $\theta (x) = \theta_T x/L$. From this definition it is clear that the superfluid fraction, $f_s$, can be expressed as:
\begin{equation}
f_s =\frac{2m}{\hbar^2}\frac{L^2}{N} \frac{\langle H_{\theta} \rangle - \langle H \rangle}{\theta_T^2} = \frac{1}{JN}\frac{\langle H_{\theta} \rangle - \langle H \rangle}{\theta^2}
\end{equation}
which is essentially the energy difference between the perturbed and the original Hamiltonian, divided by the phase squared. Note that here we used the fact that $J = \hbar^2/2ma^2$ and $\theta = \theta_T/N$, where $a$ is the typical lattice spacing. Computing $\langle H_{\theta} \rangle$ to the second order in $\theta$, we obtain the superfluid fraction  $XX$ to be:
\begin{equation}
f_s = -\frac{1}{2JN} \langle\sum_i \sigma^x_i\sigma^x_{i+1} + \sigma^y_i\sigma^y_{i+1}\rangle\\
-i \langle\sum_i \sigma^+_i\sigma^-_{i+1} - \sigma^-_i\sigma^+_{i+1}\rangle_2
\end{equation}
The second term, usually known as the superfluid current\cite{Shastry2}, vanishes for our translationally invariant 
Hamiltonian (as can easily be checked). The subscript ``2" indicates that the average is a more complicated second order one, but the exact details do not concern us here (see \cite{Shastry2}). The condition for non-zero superfluid density, $f_s >0$, now leads to
\begin{equation}
\frac{1}{2JN} \langle\sum_i \sigma^x_i\sigma^x_{i+1} + \sigma^y_i\sigma^y_{i+1}\rangle \neq 0
\label{condition}
\end{equation} 
So the supercurrent and therefore criticality exist as long as the above expectation remains finite,
which in turn implies that the condition for criticality is $\mu<J$ (so long as we are at $T=0$). When, on the other hand, $\mu\geq J$, the superfluid density identically vanishes. The critical point is therefore $\mu=J$ and the state above critical point is $|00...0\rangle$, signifying that all sites in he system are empty. Physically this is because the chemical potential, $\mu$, is too high for any particles to exist (note that the $\mu$ influences the form of the ground state and, therefore, it also affects the expectation value determining the superfluid fraction, even tough it does not appear explicitly in the equation eq. (\ref{condition})).  

Let us turn to the existence of entanglement. Since at zero temperature the state is pure it is easy to determine when it is entangled and when not. And exactly the same critical point (as for superfluidity) is identified as the borderline, $J=\mu$. We can quantify entanglement in a number of ways, and single site entropy is just one of them. This is given by $S = -\epsilon_{+}\ln\epsilon_{+} - \epsilon_{-}\ln\epsilon_{-}$,
where $\epsilon_{\pm} = (1\pm \langle \sigma^z\rangle)/2$. For $\mu>J$, $\langle \sigma^z\rangle=1$, and $S$ vanishes. Otherwise, entanglement always exists and peaks at the point when $\mu=0$. At zero temperature, therefore, superfluidity and particle number entanglement \cite{Marcelo} are fully equivalent (this now formally proves partially supported claims in \cite{V1}). Long range order, on the other hand, vanishes, but short range order does exist as discussed above. The just described phase transition at $J=\mu$ and $T=0$ is an example of a quantum BKT phase transition \cite{Sachdev}; above criticality the long range order vanishes, whereas below it suffers a polynomial drop. 

At finite temperature we can no longer assume that entanglement is quantified by $S$ since the overall state is mixed \cite{first}. We thus have to resort to the technique of entanglement witnessing. We have addressed this problem in several previous publications \cite{Brukner,BVZ,Hide} (see also \cite{Toth}). 
The result, when applied to the current scenario says that
$f_s > 1/2$ implies the existence of entanglement (this is because the correlations leading to this value of the superfluid density can only possibly be explained by assuming entangled states). Below this value of $f_s$ we still have the superfluid, but it is not clear what we can conclude about entanglement. The two phenomena are therefore at finite temperature most likely not fully equivalent, though they still are very much related. At high temperature the condition for entanglement can roughly be expressed as $\mu^2+T^2<J^2$ \cite{Brukner}. Therefore, at temperatures where superfluidity exists, we can conclude that entanglement also exists. 

Nearest neighbour entanglement can likewise be calculated, since all this requires are the knowledge of $\langle \sigma^x_i\sigma^x_{i+1}\rangle$, $\langle \sigma^z_i\sigma^z_{i+1}\rangle$ and $\langle \sigma^z_i\rangle$. The concurrence, $C$, between any nearest neighbours is then given by $C = \max \{0,|\langle \sigma^x_i\sigma^x_{i+1}\rangle|-\sqrt{(1+\langle \sigma^z_i\sigma^z_{i+1}\rangle)^2-4\langle \sigma^z_i\rangle^2}\}$.   Furthermore, $\langle \sigma^x_i\sigma^x_{i+1}\rangle = (U-\mu M)/2J$ \cite{Brukner} so that at $\mu=T=0$ we have that $C=0.04$ \cite{Gu}. Similar methods can be used to calculate nearest neighbour entanglement at finite temperature  \cite{Bortz}, but the conclusion would again be that the threshold is $J>T$ for entanglement.   

We now connect this result to our previous (continuum) investigations \cite{Anders1}. We can think of the thermal de Broglie wavelength for spins as $\lambda_{dB} = a \sqrt{J/T}$. The condition for entanglement should be that the wavelength is larger than the lattice spacing $a$ which leads us to the condition that $T<J$. Since $J=\hbar^2/2ma^2$, this implies $T < \hbar^2/2ma^2$, which is clearly consistent with our previous conclusions \cite{Anders2}. This inequality will now be applied to recent experiments.

\section{Experimental considerations} 

How would we confirm the existance of spatial entanglement experimentally? To be able to do so we need to extract the value of the nearest neighbour two point correlation function $\langle \sigma^+_i\sigma^-_{j+1}\rangle$. One way of estimating this is by interfering independent fluctuating condensates \cite{Polkovnikov}.  Let us now look at the recent remarkable experiment of Hadzibabic et al \cite{Hadzibabic} and investigate if their experiments reveal any entanglement in two dimensional Bose gases by interfering independent condensates. In their experiments the value of the chemical potential is $\mu/h = 10 KHz$, while the (average) separation between the atoms will be taken to be less than or equal to the healing length $a=0.2\mu m$ (as is true for a dilute Bose gas analysed here, see the Appendix). The temperature reached during the experiment were $100-200 nK$. They used Rubidium atoms whose mass is $87$ atomic units. Putting these numbers together we obtain that $\mu^2+(kT)^2 <J^2$, which as we showed was our criterion for the existence of entanglement. Therefore, Bose condensates exhibit field theoretic (continuous variable) entanglement (previously we have only had evidence of spin based entanglement in many body systems \cite{Ghosh,BVZ}). Furthermore, we see that the onset of entanglement is very closely related to the BKT transition. The BKT transition was measured to occur at temperatures for which the thermal de Broglie wavelength is $\lambda_T = 0.3\mu m$. The healing length, which we can think of as the effective lattice spacing, is $a=0.2\mu m$. Entanglement occurs at $a<\lambda$ which agrees with the BKT transition. This can also be obtained form the aforementioned field theoretic criterion \cite{Anders1} that $kT < J =\hbar^2/(2m a)$, where $a$ is the healing length. What is more, we can even estimate the amount of entanglement, as quantified by the single spin reduced entropy. Since $\epsilon_{\pm} = 1\pm (1-2/\pi \arccos (\mu/J))$, the (single spin) entanglement is $E \approx 0.33$. The fact that entanglement seems to occur at the point where the de Broglie wavelength becomes larger than the healing length, which itself physically represents the core size of a vortex, would suggest that entanglement may be linked to the vortex-anti vortex pairing that characterises the BKT superfluid phase. This point requires a further more in-depth study.    

\section{Conclusions}

We have seen that the traditional indicators of order in matter are, superficially speaking, not necessarily strictly related to the existence of entanglement between the underlying constituents. However, when the whole situation is properly analysed using the information given by a system's Hamiltonian we see that some rigorous relationships naturally emerge. We have found conditions for which superfluidity implies entanglement and have then go onto demonstrating that some recent experiments have already been sufficiently detailed to witness entanglement in Bose gases.
Furthermore, based on very general assumptions, it can be shown that no long range order is possible in one and
two dimensions, while it is clearly possible to have entanglement under the same circumstances. It is therefore tempting to speculate that entanglement is related to low dimensional quantum criticality. The present paper substantiates this view with an entanglement analysis of recent cold atom experiments investigating the BKT trantions. Using this entanglement for practical purposes would be the next desirable item to investigate in greater detail.  

\section{Appendix}

\subsection{The continuum limit} 

We start with the kinetic second quantised Hamiltonian:
\begin{equation}
H = \frac{\hbar^2}{2m} \int \Psi^{\dagger} (x) \nabla^2 \Psi (x) dx  
\end{equation}
where $\Psi (x)$ represents the annihilation operator for a boson.  
Suppose that bosons are localised around some lattice points (e.g. positions of fixed atoms). Then 
$\Psi (x) \approx \sum_i \phi (x-x_i) b_i$ and $\Psi^{\dagger} (x) \approx \sum_i \phi^* (x-x_i) b^{\dagger}_i$,
where $\phi (x)$ are the so called Wannier functions. Substituting this into the Hamiltonian leads to 
\begin{equation}
H = - J \sum_{ij}  b^{\dagger}_ib_j + U \sum_i n_i (n_i-1)
\end{equation}
where $J = \hbar^2/2m \int \phi^{*} (x-x_i) \nabla^2 \phi (x-x_j) dx$, $U = 4\pi\hbar^2 s/m \int dx |\phi (x)|^4$ and $s$ is the scattering length. If the $\phi$ function are well localised than only the nearest neighbour coupling will the relevant, which is what we assumed throughout the paper. The on site term is just a constant referring to the total number of atoms. If the density of atoms is low, then the $n_i^2$ term can be ignored and we obtain the $XX$ Hamiltonian analysed in the paper. The average energy per particle is, at low densities, given by $E\approx (h^2/2m) (\rho s)$. From here, the healing length, $a$, is defined by $a=1/\sqrt{\rho s}$ so that the energy is $E\approx (h^2/2ma^2)$. This leads to the following witness of entanglement that was mentioned in the text: if $\langle H\rangle < (h^2/2ma^2)$, the system is then entangled \cite{Anders1,Anders2}. This leads to the critical temperature since $\langle H\rangle$ is a function of temperature. Approximately, $\langle H\rangle\approx h^2/2m \lambda^2_T$, where $\lambda_T$ is the thermal wavelength, so the criterion for entanglement is that $\lambda_T > a$, which requires the thermal wavelength to be larger than the healing length. 

\subsection{Two dimensional XX lattice} 

True signatures of superfluidity, such as quantized vortices, can really only exist in two dimensions and higher. We will see, however, that there will no fundamental changes to our earlier one dimensional conclusions. The two dimensional $XX$ Hamiltonian is given by:
\begin{eqnarray}
H &=& \sum_i \sum_j J (\sigma^x_{i,j}\sigma^x_{i+1,j} + \sigma^y_{i,j}\sigma^y_{i+1,j}) \nonumber\\ & + & J_{\bot}(\sigma^x_{i,j}\sigma^x_{i,j+1} + \sigma^y_{i,j}\sigma^y_{i,j+1}) 
\end{eqnarray} 
We do not know how to diagonalise this Hamiltonian exactly. However, we can do reasonably well by first applying generalised (two dimensional) Jordan-Wigner transformation, and then a mean field approximation on the phases. The Jordan-Wigner transformation in two dimensions is defined as follows \cite{Fradkin}:
$\sigma^- =  e^{i\alpha_{i,j}} d_{i,j}$ and $\sigma^+ =  e^{-i\alpha_{i,j}} d_{i,j}$.
Applying the mean-field on the phases $\alpha$ we obtain the transformed Hamiltonian
of the form \cite{Fradkin}:
\begin{eqnarray}
H & = & \frac{1}{2} \sum_i\sum_j J (-1)^{i+j} (d^{\dagger}_{i,j}d_{i+1,j} - d_{i,j}d^{\dagger}_{i+1,j})\nonumber\\
& + & J_{\bot} (d^{\dagger}_{i,j}d_{i,j+1} - d_{i,j}d^{\dagger}_{i,j+1})
\end{eqnarray}
which, upon diagonalisation by the two dimensional Fourier Transform, gives \cite{2DJW}:
$H = \sum_k \Lambda (k) (\eta_k^{\dagger}\eta_k -\frac{1}{2})$,
where the eigenvalues are $\Lambda (k) = \sqrt{J^2_{\bot} \cos^2 k_y + J^2 \sin^2 k_x}$.
This now presents the (approximately) diagonalised two dimensional XX model. Let us now use the 
energy based entanglement witness \cite{Brukner} to analyse this Hamiltonian. The energy can easily be calculated from the partition function and is given by 
$U = -\frac{1}{2} \int \frac{d^2k}{(2\pi)^2} \Lambda \tanh \frac{\beta\Lambda}{2}$. 
Applying our energy based entanglement witness now reads: $|U_{\rho}|>(J+J_{\perp})/2$ implies that the state $\rho$ is entangled. Let us take the low and high temperature limit of the energy formula to draw some conclusions on entanglement. At low $T$, $\tanh \rightarrow 1$ and the (absolute value of) energy is demonstrably larger than the separable bound. The ground state in two dimensions is therefore provably entangled. At high $T$, $\tanh x \rightarrow x$ and so $|U| \rightarrow \beta (J^2+J_{\perp})^2/8$, which implies entanglement for temperatures such that 
$kT < \frac{1}{8} \frac{J^2+J_{\perp}^2}{J + J_{\perp}}$, which is very similar to the one dimensional result.  

{\it Acknowledgements:} Stimulating discussions with Janet Anders, Jacob Dunningham, Libby Heaney, Jenny Hide, Dagomir Kaszlikowski, Ian Lawrie and Wonmin Son and gratefully acknowledged.  I would also like to thank Engineering and Physical Research Council and the Royal Society Wolfson Research Merit Award for funding.

\end{document}